\documentclass[aps,pre,twocolumn,showpacs,floatfix,supperscriptaddress]{revtex4-1}
\usepackage{graphicx,bm,amsmath,dcolumn,longtable,color}
\graphicspath{{figures/}}
\newcommand{\be}{\begin{equation}}
\newcommand{\ee}{\end{equation}}
\newcommand{\la}{\langle}
\newcommand{\ra}{\rangle}

\begin{document}
\title{R\'enyi Information flow in the Ising model with single-spin dynamics}
 \author{Zehui Deng}
\affiliation{Physics Department, Beijing Normal University, Beijing 100875, China}
\author{Jinshan Wu}
\affiliation{School of Systems Science, Beijing Normal University, Beijing 100875, China}
\author{Wenan Guo}
\email{waguo@bnu.edu.cn}
\affiliation{Physics Department, Beijing Normal University, Beijing 100875, China}
\affiliation{State Key Laboratory of Theoretical Physics, Institute of Theoretical Physics, Chinese Academy of Science, Beijing 100190, China}
\date{\today}
\begin{abstract}
The $n$-index R\'enyi mutual information and transfer entropies for the 
two-dimensional kinetic Ising model with arbitrary single-spin dynamics in the
thermodynamic limit are derived as functions of ensemble averages of observables and spin-flip
probabilities. 
Cluster Monte Carlo algorithms with different dynamics from the 
single-spin dynamics are thus applicable to estimate the transfer entropies.  
By means of Monte Carlo simulations with the Wolff algorithm, we calculate the 
information flows in the Ising model with the Metropolis dynamics and the 
Glauber dynamics,  respectively.
We find that, not only the global R\'enyi transfer entropy, but also the 
pairwise R\'enyi transfer entropy peaks in the disorder phase. 
\end{abstract}
\maketitle
\section{Introduction}
Information theory has found its fruitful applications recently
\cite{matsuda,gu,melko} in the study of phase transitions and critical 
phenomenon which, traditionally, are studied by using measures based on 
two-point correlation functions.
This may not be surprising considering the concept of entropy, which was 
first used by Shannon to quantify information\cite{shannon}, has its roots in 
thermodynamics.
Mutual information (MI) has proved to be a powerful tool for 
determining the thermal and quantum phase transitions 
and their universality classes without knowledge of order parameter
\cite{melko, Amico, Cardy, Metlitski, Singh, Iaconis, inglis, Kallin}. 
(In the context of quantum critical phenomena, the classical Shannon entropy
is related to the Von Neumann entropy, and the MI is related
to the entanglement entropy\cite{Amico}.)

Besides physical systems, there are other 
complex systems with interacting agents, such as stock markets, crowd dynamics or traffic flow, also have phase-
transition-like phenomena. In such more general complex systems, there might not be a well-defined order parameter, or 
even a well-defined driven parameter as temperature to the Ising model. To predict, or even identify,   
phase-transition-like behaviors in these systems is very important, but hard. MI is therefore very useful in
studying such systems, 
e.g., Vicsek's particle swarm model\cite{Wicks}, random Boolean networks \cite{boolean} and financial 
markets\cite{harre}. 
R\'enyi entropy \cite{renyi} and corresponding mutual entropy,  as the extensions of 
the Shannon entropy and mutual entropy, also play important roles in these
methods.

On the other hand, with human civilization dives deeper and deeper into the era of big data, 
time-series data in the complex systems become more readily accessible. In principle, time-series data before and after the critical point 
should have different qualitative features. Methodologies to identify and 
predict critical points from time-series data in these systems, if established, will be an essential step of progress to research works.
Unfortunately, mutual information does not contain dynamical information. 
However, an alternative information theoretic measure, transfer entropy,
that shares some of the desired properties of mutual information but takes the information flow
into account has been introduced \cite{thomas}. 
For example, consider two Ising spins $s_1$ and $s_2$ coupled by exchange interaction. 
Let $s_1(t)$ and $s_2(t), t=1,2,\cdots,$ denote sequences of states of the two spins. 
If the state of $s_2$ has no influence  
on the transition of $s_1$, e.g., in the high temperature limit,  we have the Markov property
$p(s_1(t)\mid s_{1}(t-1))= p(s_{1}(t)\mid s_{1}(t-1), s_{2}(t-1))$
with $p$ denoting the transition probability of $s_1$ from $t-1$ to $t$. This means that there is no information 
flow from $s_2$ to $s_1$.
The deviation from this relation is thus quantified as the transfer entropy\cite{thomas}.
The transfer entropy detects the directed exchange of information between two systems
and thus might has more potential applications in the study of dynamic systems with time-series data available
\cite{price, marinazzo}.

For a complex dynamic system, it is known that information flow between elements always peak in an intermediate order 
regime. However, the peak may not coincide with phase transition. 
It was recently conjectured that, by contrast, information flow 
in such systems generally may peak strictly at the disordered side of a phase
transition \cite{barnett}. This conjecture was verified for the ferromagnetic
two-dimensional(2D) kinetic Ising model with the Glauber dynamics\cite{glauber},
in which a global transfer entropy measure attains a maximum in the disordered
phase. However, a pairwise transfer entropy measure does not show such a 
maximum in the disordered phase\cite{barnett}.
The numerically observed peak of global transfer entropy at the disorder side can be 
practically very valuable. In a stock market, ordered phase, where lots of stocks move in the same direction, in a sense corresponds to a large 
bubble or a big crash. Peaks in the disorder side implies that it might be used as an indicator of a critical region in the near future before stocks 
in the market really start to act in the same direction.

The MI and information flow discussed in Ref. \cite{barnett} are based on the 
Shannon entropy. It is natural to extend the theory to include the general 
R\'enyi entropy\cite{renyi} and to verify other single-spin dynamics than the Glauber dynamics. 
In present work, we define the R\'enyi pairwise and global
MI measures and the corresponding transfer entropy measures. 
For the 2D kinetic Ising model with general single-spin dynamics, 
these measures are derived as functions of ensemble averages of observables, including
those related to the single-spin flipping probabilities, in the thermodynamic limit.
We further numerically calculate these measures by using Monte Carlo 
simulations with the Wolff algorithm\cite{wolff}. 
We find that the Shannon transfer entropies for the Ising model with the Metropolis 
dynamics \cite{metropolis} behave similarly as those for the Glauber dynamics.
The R\'enyi pairwise and global MI measures are also found to have similar 
behaviors as the Shannon counterparts for both dynamics. 
However, the R\'enyi pairwise and global transfer entropies 
show different behaviors from the Shannon counterparts. The most 
evident difference is that the R\'enyi pairwise transfer information measure 
peaks in the disordered phase, which is absent for the Shannon pairwise
information flow.

The paper is organized as follows:
In Sec II, we define and derive the R\'enyi entropy based MI and flow
measures in the thermodynamic limit. In Sec III, we calculate the measures 
for the 2D kinetic Ising model with Glauber and Metropolis dynamics.
We conclude in Sec IV.

\section{R\'enyi mutual information and flow}
We consider the ferromagnetic 2D Ising model on the square lattice with periodic boundary conditions.
The Hamiltonian is given by
\begin{equation}
{\cal H} ({\bf S})= -J\sum_{\langle{i,j}\rangle}S_{i} S_{j},
\end {equation}
where ${\bf S}= (S_{1},..., S_{N}), S_{i}\in \{+1,-1\}$, denotes the spin 
configuration and $\langle{i,j}\rangle$ the nearest neighbors.  
$J=1$ sets the energy unit. The Boltzmann-Gibbs probability of a configuration
${\bf S}$ is
\begin{equation}
{\bf P}({\bf S})=\frac{1}{Z} {\rm e}^{-\beta {\cal H}({\bf S})},
\end{equation}
where $\beta=1/T$ is the inverse temperature with  the Boltzmann constant 
$k_{\rm B}=1$, and $Z=\sum_{\bf S} {\rm e}^{-\beta {\cal H}({\bf S})}$ is the 
partition function.

The model is largely solved in the thermodynamic limit
\cite{onsager, yang, mccoy}. We quote the main exact results here for later
use:\\
The critical inverse temperature  
\be
\beta_{\rm c}=\frac{1}{T_{\rm c}}=\frac{1}{2}\log(1+\sqrt{2}),
\ee 
the magnetization
\begin{eqnarray} 
m = \{ \begin{array}{l l} 
\pm(1-\sinh^{-4}{2\beta})^{\frac{1}{8}}, & \quad  T< T_{c}; \\
0,                                       & \quad T \geq T_{c}, 
\label{m}
\end{array}
\end{eqnarray}
the free energy per site 
\be 
-2\beta f=\log(2\cosh^{2}
{2\beta})+\frac{2}{\pi}\int^{\pi/2}_{0}\log(1+\sqrt{1-\kappa^{2}\sin^{2}\theta})d\theta,
\label{fren}
\ee
and the internal energy per site 
\be u=-\coth2\beta[1+\frac{2}{\pi}(\kappa\sinh2\beta-1)\int^{\pi/2}_{0}\frac{d\theta}{\sqrt{1-\kappa^{2}\sin^{2}\theta}}],
\label{u}
\ee 
where $\kappa=\frac{2\sinh{2\beta}}{\cosh^{2}{2\beta}}$. 

Mutual information between random variables is the essential 
information-theoretic quantity, which can be framed in terms of statistical 
dependence. Based on the Shannon entropy $H(X)$ of a random variable $X$,
the mutual information $I(X:Y\mid Z)$ between two random variables $X$ and $Y$, 
optionally conditional on a third variable $Z$, is defined as
\be\begin{split} 
I(X:Y\mid Z)& \equiv H(X\mid Z)-H(X\mid Y,Z),
\end{split}\ee
which is equivalent to
\be
I(X:Y\mid Z)
= H(X\mid Z)+H(Y\mid Z)-H(X,Y\mid Z).
\ee

Barnett {\it et al.} \cite{barnett} thus define the pairwise MI measure 
\be \label{ipw}
I_{\rm pw}=\frac{1}{2N}\sum_{\la i, j \ra} I(S_i:S_j)
=\frac{1}{2N} \sum_{\la i, j \ra}(2 H(S_i)-H(S_i,S_j)),
\ee
and the global MI measure as the multi-information 
\be \label{igl}
I_{\rm gl}=\sum_i H(S_i)-H({\bf S}).
\ee 
The parametric family of entropies so called R\'enyi 
entropy were introduced by Alfred R\'enyi 
as a mathematical generalization of the Shannon entropy. 
The definition of R\'enyi's entropy of index $n$ is given by\cite{renyi}
\begin{equation}\label{hn}
H_{n}(X)=\frac{1}{1-n}\log(\sum_{i\in{X}}p^{n}_{i}),
\end{equation}
where $X$ represents a random variable and $p_{i}$ is the probability of 
outcome $i\in{X}$. 
Shannon entropy is the special case at the limit $n\rightarrow 1$.  

There are alternatives to define the R\'enyi MI
$I_{n}(X:Y)$ between two random variables $X,Y$ 
\cite{Jose}, e.g., $I_{n}(X:Y)\equiv H_{n}(X)-H_{n}(X\mid Y)$, or,
$I_{n}(X:Y)\equiv H_{n}(X)+H_{n}(Y)-H_{n}(X,Y).$
Following Iaconis {\it et al.} \cite{Iaconis},
we adopt the latter and 
extend $I_{\rm pw}$ to the R\'enyi pairwise MI measure $I^{\rm R}_{\rm pw}$,
and $I_{\rm gl}$ to the R\'enyi global MI measure $I^{\rm R}_{\rm gl}$
by replacing the Shannon entropy $H$ to the R\'enyi entropy $H_n$ in
Eq. (\ref{ipw}) and (\ref{igl}), respectively.

Following Barnett {\it et al.}\cite{barnett}, we express them in the 
thermodynamic limit  
\begin{equation}\label{rmpw}
I^{\rm R}_{\rm pw}= \frac{2}{1-n}\log(\sum_{\sigma}p^{n}_{\sigma})-\frac{1}{1-n}\log(\sum_{\sigma,\sigma^{\prime}}p^{n}_{\sigma\sigma^{\prime}})
\end{equation}
and
\begin{equation}\label{rmgl}
\frac{1}{N}I^{\rm R}_{\rm gl}= \frac{1}{1-n}\log(\sum_{\sigma}p^{n}_{\sigma})+\frac{n\beta}{1-n}(f(T/n)-f(T)),
\end{equation}
where the sums are over $\sigma,\sigma^{\prime}=\pm1$, with
\begin{equation}\label{pp}
p_{\sigma}=\frac{1}{2}(1+\sigma{m}),\quad p_{\sigma\sigma^{\prime}}=\frac{1}{4}[1+(\sigma+\sigma^{\prime})m-\frac{1}{2}\sigma\sigma^{\prime}\it{u}],
\end{equation}
and $n$ is the index of the R\'enyi entropy, $m, f$ and $u$ is the 
magnetization, the free energy persite and the internal energy persite, 
respectively.
Note that for $T<T_{c}$, 
the sign of the magnetization $m$ does not affect these two and 
any subsequent quantities, which is to say that the information measures are 
invariant under symmetry breaking.

$I^{\rm R}_{\rm pw}$ and $I^{\rm R}_{\rm gl}$, at the thermodynamic limit, 
can be computed directly by substituting the 
exact results as in Eq. (\ref{m}-{u}) into their analytic expressions (\ref{rmpw}) and (\ref{rmgl}). 
Also we note that the second derivative of $I^{\rm R}_{\rm gl}$ has 
singular points at $T_{\rm c}$ and $nT_{\rm c}$ due to the singular behavior of the
free energy.

To study the information flow between stationary stochastic processes 
$X(t)$ and $Y(t)$, the transfer entropy $T_{Y\rightarrow X} \equiv
I(X(t):Y^{(l)}(t)\mid X^{(l)}(t))$ with $l$-length history
is useful. Here $X^{(l)} \equiv X(t-1), \dots, X(t-l)$.
Barnett {\it et al.} \cite{barnett} considered the $l=1$ history pairwise 
transfer entropy measure and global transfer entropy measure based on the 
Shannon entropy:
\begin{eqnarray}\label{tpw1}
T_{\rm pw}&=&\frac{1}{2N} \sum_{\la i,j \ra} T_{S_j \rightarrow S_i} \\ \nonumber
      &=&\frac{1}{2N} \sum_{\la i, j \ra} (H(S_{i}(t)\mid S_{i}(t-1)) \\ \nonumber
      && -H(S_{i}(t)\mid S_{i}(t-1),S_{j}(t-1))),
\end{eqnarray}
and
\be\label{tgl1}
T_{\rm gl}= \sum_i (H(S_{i}(t)\mid S_{i}(t-1)) - H(S_{i}(t)\mid {\bf S}(t-1))),
\ee
where $S_{i}(t)$ denotes the spin $i$ at time $t$, $S_{j}(t-1)$ 
represents the neighboring spin $j$ at time $t-1$, 
$ H(S_{i}(t)\mid S_{i}(t-1))$ is the Shannon entropy of $S_{i}(t)$ conditional 
on $S_{i}(t-1)$ and similarly for the others.
Starting from these definitions, these measures are calculated for an arbitrary 
single-spin dynamics of the Ising model in the thermodynamic limit\cite{barnett}, where 
exact results (Eq. (\ref{m})-(\ref{u})) in the 
thermodynamic limit are used. 
For the sake of completeness, we quote the their results as follows: 
\begin{equation}\label{tpw}
N\it{T}_{\rm pw}= -q\sum_{\sigma}\log\frac{q}{p_{\sigma}}+\sum_{\sigma^{\prime}}q_{\sigma^{\prime}}\sum_{\sigma}\log\frac{q_{\sigma^{\prime}}}{p_{\sigma\sigma^{\prime}}},
\end{equation}
and
\begin{equation}\label{tgl}
NT_{\rm gl}=-q\sum_{\sigma}\log\frac{q}{p_{\sigma}}+\langle{P_{i}({\bf S})\log{P_{i}({\bf S})}}\rangle,
\end{equation}
where
\begin{equation}\label{qq}
q=\frac{1}{2}\langle{P_{i}({\bf S})}\rangle,\quad q_{\sigma^{\prime}}=\frac{1}{4}(\langle{P_{i}( {\bf S})}\rangle+\sigma^{\prime}\langle{S_{j}P_{i}({\bf S})}\rangle),
\end{equation}
with $i, j$ arbitrary nearest neighbors and $\langle{S_{j}P_{i}({\bf S})}\rangle\equiv 0$ for $T\geq T_{c}$; 
$P_i({\bf S)}$ is the flipping probability of spin $S_i$ in a given spin 
configuration $\bf{S}$\cite{barnett}, which describes any 
single-spin process as long as this process satisfies the detailed balance.
It is important to notice that the MI measures are independent
of the dynamics, while the transfer entropy measures do depend on the 
dynamics.

We can also generalize the pairwise and global transfer (Shannon) entropy 
measures to the R\'enyi pairwise and R\'enyi global transfer entropy 
measures.  For two stationary stochastic processes $X(t)$ and $Y(t)$, 
we define the $l=1$-length history R\'enyi transfer entropy 
\be 
T^R_{Y\rightarrow X} \equiv 
H_{n}(X(t)|X(t-1)-H_{n}(X(t)\mid X(t-1), Y(t-1)),
\ee
which reduces to the Shanon transfer entropy $T_{Y \to X}$ at the limit $n \to 1$.
The R\'enyi pairwise and R\'enyi global transfer entropy
measures are thus defined by replacing $H$ to $H_n$ in Eq. (\ref{tpw1}) and
(\ref{tgl1}), respectively.
The expressions, at thermodynamic limit, are
found to be
\begin{multline}\label{trpw}
T^{\rm R}_{\rm pw}= \frac{1}{1-n}\sum_{\sigma}p_{\sigma}\log[(1-\frac{q}{Np_{\sigma}})^{n}+(\frac{q}{Np_{\sigma}})^{n}]\\
-\frac{1}{1-n}\sum_{\sigma, \sigma^{\prime}}p_{\sigma\sigma^{\prime}}\log[(1-\frac{q_{\sigma^{\prime}}}{Np_{\sigma\sigma^{\prime}}})^{n}+(\frac{q_{\sigma^{\prime}}}{Np_{\sigma\sigma^{\prime}}})^{n}]\\
\end{multline}
and
\begin{multline}\label{trgl}
T^{\rm R}_{\rm gl}= \frac{1}{1-n}\sum_{\sigma}p_{\sigma}\log[(1-\frac{q}{Np_{\sigma}})^{n}+(\frac{q}{Np_{\sigma}})^{n}]\\
-\frac{1}{1-n}\langle{\log((1-\frac{P_{i}(\bf S)}{N})^{n}+(\frac{P_{i}(\bf S)}{N})^{n})}\rangle,
\end{multline}
respectively. Here, $p_{\sigma}$, $p_{\sigma\sigma^{\prime}}$, $q$ and $q_{\sigma^{\prime}}$ are defined in Eq.(\ref{pp}) and (\ref{qq}).

For large system $N\rightarrow \infty$, we obtain 
the index $n=2$ R\'enyi $T^{\rm R}_{\rm pw}$ and $T^{\rm R}_{\rm gl}$, 
by  applying Taylor expansion, to the order ${\bf O}(\frac{1}{N^{3}})$, 
respectively:
\begin{equation}\label{trpwx}
\begin{split}
T^{\rm R}_{\rm pw} & = -\sum_{\sigma}p_{\sigma}(-2\frac{q}{Np_{\sigma}}+\frac{4}{3}\frac{q^{3}}{N^{3}p^{3}_{\sigma}})+\sum_{\sigma, \sigma^{\prime}}p_{\sigma\sigma^{\prime}}(-2\frac{q_{\sigma^{\prime}}}{Np_{\sigma\sigma^{\prime}}}\\
& +\frac{4}{3}\frac{q^{3}_{\sigma^{\prime}}}{N^{3}p^{3}_{\sigma\sigma^{\prime}}})+{\bf O}(\frac{1}{N^{4}})\\
&=-\frac{4}{3N^{3}}(\sum_{\sigma}\frac{q^{3}}{p^{2}_{\sigma}}-\sum_{\sigma, \sigma^{\prime}}\frac{q^{3}_{\sigma^{\prime}}}{p^{2}_{\sigma\sigma^{\prime}}})+{\bf O}(\frac{1}{N^{4}})
\end{split}
\end {equation}
and
\begin{equation}\label{trgly}
\begin{split}
T^{\rm R}_{\rm gl} & = -\sum_{\sigma}p_{\sigma}(-2\frac{q}{Np_{\sigma}}+\frac{4}{3}\frac{q^{3}}{N^{3}p^{3}_{\sigma}})+\langle{-2\frac{P_{i}({\bf S})}{N}+\frac{4}{3}\frac{P^{3}_{i}({\bf S})}{N^{3}}}\rangle\\
&+{\bf O}(\frac{1}{N^{4}})\\
&=-\frac{4}{3N^{3}}(\sum_{\sigma}\frac{q^{3}}{p^{2}_{\sigma}}-\langle{P^{3}_{i}({\bf S})}\rangle)+{\bf O}(\frac{1}{N^{4}}).
\end{split} 
\end {equation}
One great advantage of these two formulas is that they are expressed in terms of ensemble averages of observables, 
based only on the Boltzmann-Gibbs distribution. The nature of the transfer entropies
which are sensitive to the update scheme is represented in ensemble averages of quantities like
$\langle{P_{i}({\bf S})}\rangle$ and $\langle{S_{j}P_{i}({\bf S})}\rangle)$, which can
be calculated by using efficient MC method with dynamics other than the single-spin
dynamics involved in the kinetic model, given that the  update probability $P_i({\bf S})$ is
specified. Simulation results of
these two quantities for two dynamics are presented in Section \ref{IIIA}.

\section{Numerical results}
The Metropolis algorithm is the first MC algorithm to simulating
lattice models\cite{metropolis}.
The underlying discrete-time Metropolis spin-flip dynamics is defined as 
follows: 
at each time step, an arbitrary spin $i$ is chosen randomly.
Consider the energy difference between the state that spin $i$ is flipped and 
the original state: 
$\Delta{E}_i=2 s_{i}\sum_{j\in\nu(i)}s_{j}$,
$\nu(i)$ denotes the nearest neighbors of spin $i$. 
The spin-flipped state will be accepted with 
probability $1$, if $\Delta{E}_i \le {0}$;  
otherwise, the state will be accepted with the probability
\begin{equation}
P_{i}({\bf S})={\rm e}^{-\Delta{E}_i /T }\label{com_map}.
\end{equation}
The discrete-time Glauber spin-flip dynamics\cite{glauber} is slightly 
different from the Metropolis dynamics: 
The randomly chosen spin $i$ flips with the probability 
\be
P_i({\bf S})=[1+{\rm e} ^{\Delta{E}_i/ T}]^{-1}.
\label{prb_glauber}
\ee
These processes satisfy detailed balance. 

Since not every term of the transfer entropies has analytic expression, we 
make use of MC method to obtain their behavior. 
The Wolff cluster algorithm\cite{wolff} is used to generate microscopic states.
The ensemble average of an observable is calculated as means in the samples. 
This algorithm is much
more efficient than other single-spin flip algorithms, such as the Metropolis 
algorithm and its variations. In particular, it suppresses critical slowing 
down.  In our simulations, typically $10^5$ samples are used to obtain ensemble 
averages and statistical errors after equilibrating the systems. It is worthy
to note that we do not study the transfer entropies of the kinetic Ising model 
with the dynamics of the Wolff algorithm. Instead, 
the Wolff MC method is used to calculate the information flows,  
according to Eqs. (\ref{tpw}), (\ref{tgl}), (\ref{trpwx}), and (\ref{trgly})), 
in the kinetic Ising model with the Metropolis and the Glauber dynamics, 
respectively.

\subsection{Shannon entropy based information flow for the Metropolis
dynamics}
\label{IIIA}

To further verify the conjecture raised in \cite{barnett} that MI flows peak 
in the disordered phase, we study the MI 
flows in the Ising model with the Metropolis dynamics.

\begin{figure}[htbp]
\includegraphics[scale=0.8]{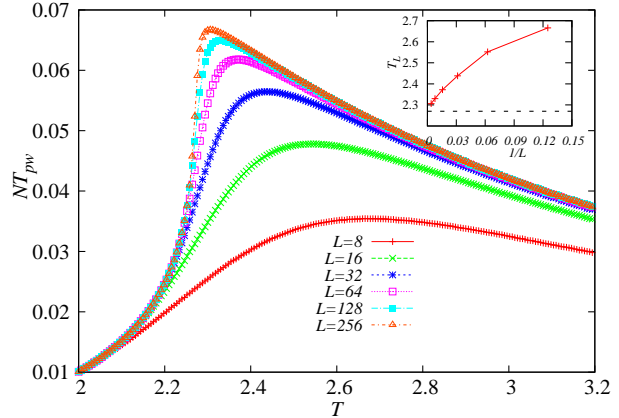}
\caption{(color online) Plot of $NT_{\rm pw}$ for the Metropolis dynamics 
against temperature for several system sizes. 
The statistical errors are much smaller than the symbol sizes. The inset shows the maximum of $T_{\rm pw}$ as 
a function of $1/L$, in which the horizontal dashed line indicates $T_c=2.2692$.}
\label{tebpw1}
\end{figure}

\begin{figure}[htbp]
\includegraphics[scale=0.75]{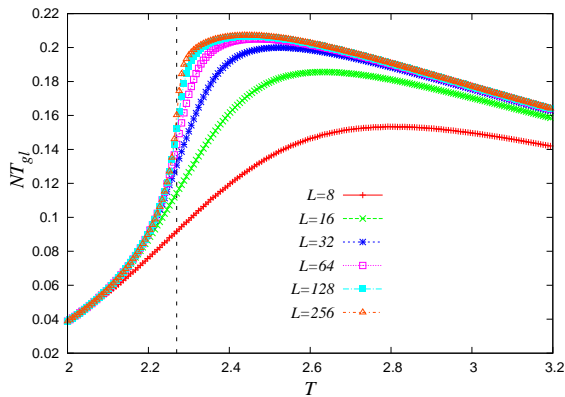}
\caption{(color online) Plot of $NT_{\rm gl}$ for the Metropolis dynamics against temperature for several system sizes. The dashed vertical line
indicates $T_c=2.2692$. The statistical errors are much
smaller than the symbol sizes.}
\label{tebgl1}
\end{figure}

We simulate the Ising model on the square lattice of 
size $N=L\times L$ for $L=8,16,32,64,128,256$.
Figure \ref{tebpw1} and \ref{tebgl1} shows $T_{\rm pw}$ and $T_{\rm gl}$ as 
functions of temperature $T$ and linear size $L$ for the Metropolis dynamics, 
respectively. 
The results are very similar to those found for the Glauber dynamics in 
Ref.\cite{barnett}. 
As the system size grows, the finite-size effects are reducing.
Kinks turn to appear in the $T_{\rm pw}$ and $T_{\rm gl}$ versus 
temperature curves at the exactly known critical point $T_{c}\approx2.2692$. 
In the inset of Fig.\ref{tebpw1}, we show the maximum of $T_{\rm pw}$ as 
a function of $1/L$, which converges to the known critical point $T_c$ very well.
This quantity can thus be used to determine the critical point of other 
systems without knowledge of the analytical solution.    
However, there are humps in the $T_{\rm gl}$ curves. The maximum 
keeps sitting in the disorder region when $N\to \infty$.
Compared with Ref. \cite{barnett} in which $T_{\rm gl}$ is found max at 
$T=2.354\pm0.003$ for the Glauber dynamics, $T_{\rm gl}$ has a 
maximum at $T=2.44\pm0.01$. 

The conclusion is the same as Barnett {\it et al.} \cite{barnett}, namely, 
$T_{\rm pw}$ peaks at $T_{\rm c}$ while $T_{\rm gl}$ has a maximum at the disordered phase.
Similar results on a measure related to $T_{\rm pw}$ have been 
obtained by direct simulating the Ising model with the Metropolis dynamics and Glauber dynamics \cite{marinazzo}.

\begin{figure}[htbp]
\includegraphics[scale=0.65]{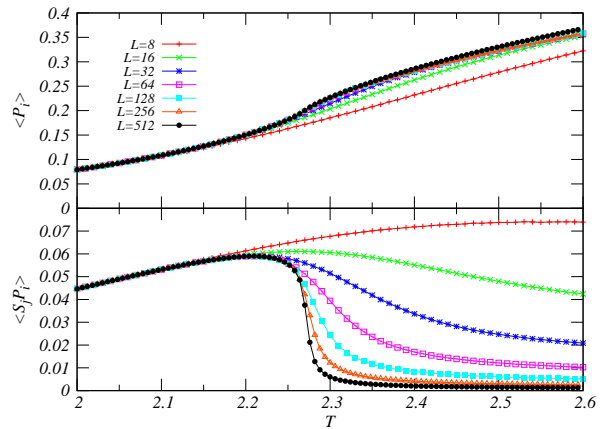}
\caption{(color online) $\left<P_i \right>$ (upper panel) and $\left<S_j P_i\right>$ (lower panel) plotted against temperature for the 2D Ising model
with the Metropolis dynamics. The statistical errors are much smaller than the symbol sizes.
}
\label{pspm}
\end{figure}

\begin{figure}[htbp]
\includegraphics[scale=0.65]{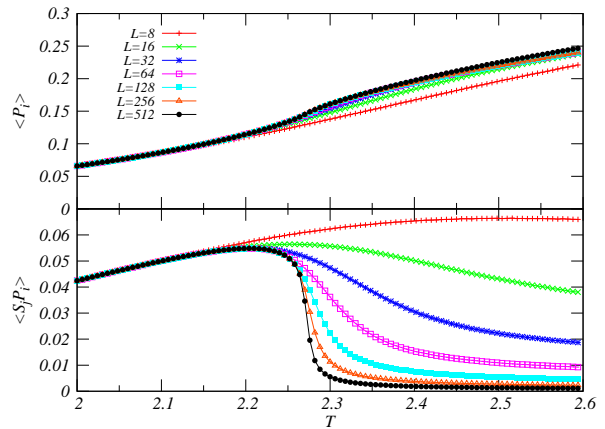}
\caption{(color online) $\left<P_i \right>$ (upper panel) and $\left<S_j P_i\right>$ (lower panel) plotted against temperature for the 2D Ising model
with the Glauber dynamics. The statistical errors are much smaller than the symbol sizes.
}
\label{pspg}
\end{figure}

It is worthy to mention that, in Ref.\cite{barnett}, the authors stressed that 
the dynamics of the spin updating in their MC algorithm is necessarily to be 
the same as the dynamics under discussing. By contrast, we come to a 
conclusion that it is irrelevant to use which MC algorithm or dynamics to 
update configurations in the simulations, as far as the MC means is equal
to the ensemble averages. 
This is because that the entropy flows have been expressed as ensemble 
averages of observables in the equilibrated system 
(see Eq. (\ref{tpw}), (\ref{tgl}), (\ref{trpwx}), and (\ref{trgly})).
There's no need to extract the time series of the entropies 
appearing in the definitions of the flows.
For example, $\left<P_i \right>$ and $\left<S_j P_i\right>$ are two essential quantities related to the specific dynamics
in above expressions, which can be determined by MC simulations
with spin update algorithms different from the dynamics studied, as far as the Boltzmann-Gibbs distribution are realized by the simulations. 
Figure \ref{pspm} and \ref{pspg} show these two quantities as functions of temperature for the kinetic Ising model with the Metropolis dynamics and with the Glauber dynamics,
respectively. The results are obtained by MC simulations with the Wolff 
algorithm. It is seen that singularities develop in the two quantities closing to the critical point when system size turns
large. We have verified this conclusion by 
repeating Barnett's results using the Wolff algorithm (not shown here).

\subsection{R\'enyi entropy based mutual information
and flow for the Glauber and Metropolis dynamics}
\label{renyi}

We now study the generalized index 2 R\'enyi  MI measures and transfer
entropy measures for the Ising
model with the Metropolis dynamics and Glauber dynamics, respectively.  

\begin{figure}[htbp]
\includegraphics[scale=0.65]{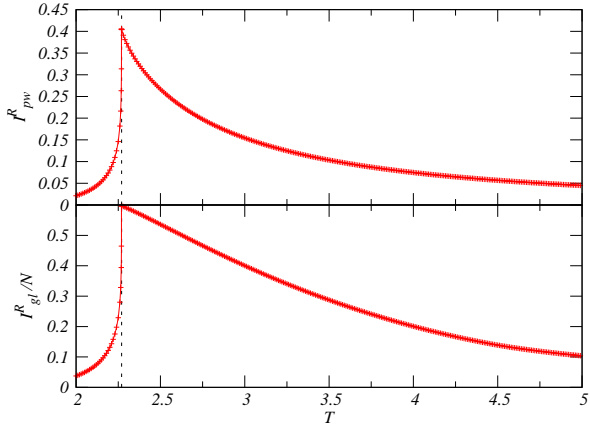}
\caption{(color online) Plot of $I^{\rm R}_{\rm pw}$ (upper panel) and
$I^{\rm R}_{\rm gl}/N$ (lower panel) against temperature.}
\label{jrtmbgl}
\end{figure}

The R\'enyi MI measures do not depend on the dynamics, thus
can be calculated analytically. By substituting the exact $m$, $f$ 
and $u$ into Eq.(\ref{rmpw}) and Eq.(\ref{rmgl}), we obtain the results, which are
plotted against temperature in Fig.\ref{jrtmbgl}. 
As expected, the R\'enyi pairwise MI and global MI
bear singularities at the exactly known critical point. 
We also expect singular behavior of $I^{\rm R}_{\rm gl}$ at $2T_{\rm c}$ due to
the singularity in the free energy (see Eq. (\ref{rmgl})). 
Such singularity is not visible directly in Fig. \ref{jrtmbgl}(lower panel), 
but should appear as a logarithmic divergence 
in the second order derivative.

By contrast, the R\'enyi pairwise MI flow $T^{\rm R}_{\rm pw}$ and global MI flow
$T^{\rm R}_{\rm gl}$ depend on the dynamics of the kinetic Ising model.
MC simulations with the Wolff algorithm are used to calculate the two 
measures.
According to (\ref{trpwx}) 
and (\ref{trgly}), the leading terms in $T^{\rm R}_{\rm pw}$ and $T^{\rm R}_{\rm gl}$ scale as 
$1/N^3$. 
We therefore evaluate $N^{3}T^{\rm R}_{\rm pw}$ and  $N^{3}T^{\rm R}_{\rm gl}$.

\begin{figure}[htbp]
\includegraphics[scale=0.65]{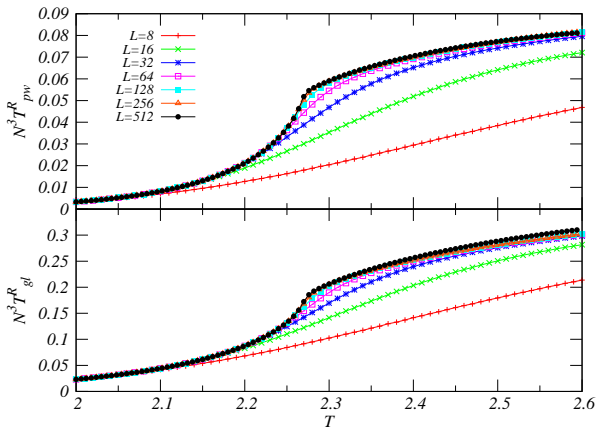}
\caption{(color online)  $N^{3}T^{\rm R}_{\rm pw}$(upper panel) and $N^{3}T^{\rm R}_{\rm gl}$(lower panel) plotted against temperature for the 2D Ising 
model with Metropolis dynamics for several system sizes $L$. 
The statistical errors are much smaller than the symbol sizes.
}
\label{jrtebgl}
\end{figure}

\begin{figure}[htbp]
\begin{center}
\includegraphics[scale=0.65]{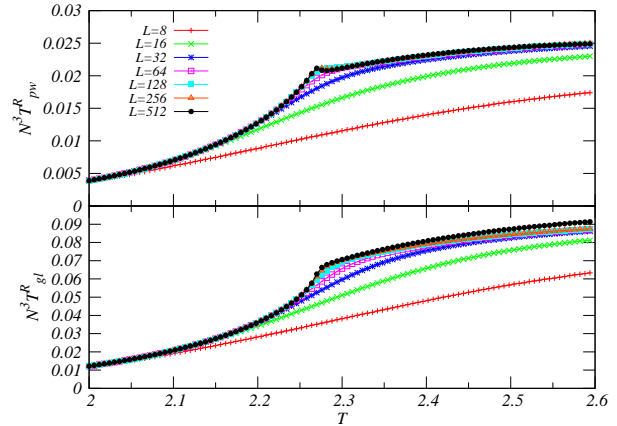}
\end{center}
\caption{(color online) $N^3 T^{\rm R}_{\rm pw}$ (upper panel)  and $N^{3}T^{\rm R}_{\rm gl}$ (lower panel) as a function of temperature for the 2D Ising model
with the Glauber dynamics for several system sizes.
The statistical errors are much smaller than the symbol sizes.}
\label{jrtebgl1}
\end{figure}

Figure \ref{jrtebgl} illustrates the 
R\'enyi transfer entropy $T^{\rm R}_{\rm pw}$ and global transfer entropy $T^{\rm R}_{\rm gl}$ 
for the Metropolis dynamics as functions of temperature for system sizes $L=8, 16, 32, 64, 128, 256, 512$, 
while  Figure \ref{jrtebgl1} shows those for the Glauber dynamics.
The two dynamic processes have similar behavior in $T^{\rm R}_{\rm pw}$ and 
$T^{\rm R}_{\rm gl}$:  as system size turns large, they all develop a kink 
around $T_{\rm c}$; the curve for each size has a hump in the disorder region.
For R\'enyi pairwise transfer entropy, the maximum point in the curve of the largest size $L=512$ is 
at $T=2.89\pm0.05$ for the Metropolis dynamics and $T=2.70\pm 0.05$ for the Glauber dynamics, respectively; 
For R\'enyi global transfer entropy, the maximum point in the curve of the largest size $L=512$ is at $T=3.21\pm0.05$ 
and $T=2.93\pm0.05$ for the two dynamics, respectively. All these curves are rather flat around the maximum points, 
which lead to large errors in the estimates of maximum points. 

All these measures peak in the disordered regime, 
regardless the type of the single-spin dynamics, is remarkable. 
In particular, the behavior of the R\'enyi global transfer entropy 
$T^{\rm R}_{\rm gl}$ is similar to the Shannon global transfer entropy 
$T_{\rm gl}$ for the Glauber dynamics, however, the 
R\'enyi pairwise transfer entropy $T^{\rm R}_{\rm pw}$ shows a completely 
different behavior from the Shannon $T_{\rm pw}$, namely, $T^{\rm R}_{\rm pw}$ 
peaks in the disordered region, while $T_{\rm pw}$ does not.

\section{Conclusions and discussion}

We have extended the Shannon pairwise, global MI and 
$l=1$ history transfer entropies to the R\'enyi counterparts. 
Expressions related to thermodynamic quantities and ensemble
averages of dynamic probability are derived in the thermodynamic limit for the 
2D kinetic Ising model with arbitrary single-spin dynamics.
Cluster Monte Carlo algorithms with different dynamics from the 
single-spin dynamics are thus applicable to estimate the transfer entropies.
As a result, much larger system sizes and numerical accuracy can be reached
in simulations.
  
By using Wolff cluster Monte Carlo simulations, we have calculated the transfer entropies for both the Shannon and the R\'neyi entropy, 
for the kinetic Ising model with the Glauber and the Metropolis dynamics. 

The Shannon global transfer entropy is shown to has a maximum point in 
the disordered regime for the Metropolis dynamics, similar to that found \cite{barnett} for 
the Glauber dynamics. Also, the Shannon pairwise transfer entropy for the
Metropolis dynamics behaves similarly as that for the Glauber dynamics \cite{barnett}:
$T_{\rm pw}$ peaks at $T_{C}$, but does not max in the disordered regime.

For the R\'enyi transfer entropies with index $2$, we have found that, in 
additional to the global transfer entropy $T_{\rm gl}$, the R\'enyi pairwise 
transfer entropy $T^{\rm R}_{\rm pw}$ 
peaks in the disordered phase for both the Metropolis and the Glauber
dynamics. This is different from the behavior of the Shannon pairwise transfer 
entropies.

$T_{\rm gl}$ is regarded as measure of collective information transfer
\cite{Lizier}, capturing both pairwise and higher-order (multivariate) 
correlations of a site. Its peak is interpreted \cite{barnett} in terms of 
conflicting tendencies amongst these components as the level of disorder in 
the system increases when the system is further away from the phase 
transition. This might also
explain the postcritical peak in our R\'enyi global transfer entropy 
$T^{\rm R}_{\rm gl}$. 
However, we don't have an intuitive explanation for the postcritical peak in 
our R\'enyi pairwise transfer entropy $T^{\rm R}_{\rm pw}$ which is absent in the 
Shannon counterpart.  Further investigation is required.

{\bf Acknowledgment}
This work is supported by the National Science Foundation of China
(NSFC) under Grant 11175018 (Guo) and 11205014 (Wu).

\end{document}